\pdfoutput=1
\pdfoutput=1
\pdfoutput=1
\pdfoutput=1
\pdfoutput=1
\documentclass[conference,10pt]{IEEEtran}
\makeatletter

\newcommand{\Rmnum}[1]{\expandafter\@slowromancap\romannumeral #1@}

\makeatother
\IEEEoverridecommandlockouts
\usepackage{amsmath,bm}
\usepackage{amssymb}
\usepackage{extarrows}
\usepackage{algorithm}
\usepackage{algpseudocode}
\usepackage{algorithmicx}
\usepackage{graphicx} %use graph format
\usepackage{epstopdf}
\usepackage{amsthm}
\usepackage{enumerate}
\usepackage[T1]{fontenc}
\usepackage[numbers,sort&compress]{natbib}
\usepackage{commath}
\usepackage{multirow}
\usepackage{siunitx}
\usepackage{cases}
\usepackage{svg}
\usepackage{graphicx}
\usepackage{float}
\usepackage{amsfonts} %% <- also included by amssymb
\DeclareMathSymbol{\shortminus}{\mathbin}{AMSa}{"39}
\usepackage{gensymb}

\algrenewcommand\algorithmicrequire{\textbf{Input: }}
\algrenewcommand\algorithmicensure{\textbf{Output: }}
\algnewcommand\algorithmicInit{\textbf{Init: }}

\renewcommand{\algorithmicrequire}{ \textbf{Input:}} %Use Input in the format of Algorithm
\renewcommand{\algorithmicensure}{ \textbf{Output:}} %Use Output in the format of Algorithm
\theoremstyle{plain}

\newtheorem{remark}{Remark}

\begin{document}
%--------------------------------------------
%command file for *.tex files
%--------------------------------------------
% \newcommand{\fs}{\hspace{0.07in}}
\newcommand{\bs}{\hspace{-0.1in}}
\newcommand{\re}{{\rm Re} \, }
\newcommand{\e}{{\rm E} \, }
\newcommand{\p}{{\rm P} \, }
\newcommand{\cn}{{\cal CN} \, }
\newcommand{\n}{{\cal N} \, }
\newcommand{\ba}{\begin{array}}
\newcommand{\ea}{\end{array}}
\newcommand{\be}{\begin{displaymath}}
\newcommand{\ee}{\end{displaymath}}
\newcommand{\ben}{\begin{equation}}  
\newcommand{\een}{\end{equation}}
\newcommand{\bea}{\begin{equation}\begin{aligned}}
\newcommand{\eea}{\end{aligned}\end{equation}}      
\newcommand{\bena}{\begin{eqnarray}}
\newcommand{\eena}{\end{eqnarray}}
\newcommand{\beqa}{\begin{eqnarray*}}
\newcommand{\enqa}{\end{eqnarray*}}
\newcommand{\f}{\frac}
\newcommand{\bc}{\begin{center}}
\newcommand{\ec}{\end{center}}
\newcommand{\bi}{\begin{itemize}}
\newcommand{\ei}{\end{itemize}}
\newcommand{\benu}{\begin{enumerate}}
\newcommand{\eenu}{\end{enumerate}}
\newcommand{\bdes}{\begin{description}}
\newcommand{\edes}{\end{description}}
\newcommand{\bt}{\begin{tabular}}
\newcommand{\et}{\end{tabular}}
\newcommand{\vs}{\vspace}
\newcommand{\hs}{\hspace}
\newcommand{\sort}{\rm sort \,}

\newcommand \thetabf{{\mbox{\boldmath$\theta$\unboldmath}}}
\newcommand{\Phibf}{\mbox{${\bf \Phi}$}}
\newcommand{\Psibf}{\mbox{${\bf \Psi}$}}
\newcommand \alphabf{\mbox{\boldmath$\alpha$\unboldmath}}
\newcommand \betabf{\mbox{\boldmath$\beta$\unboldmath}}
\newcommand \gammabf{\mbox{\boldmath$\gamma$\unboldmath}}
\newcommand \deltabf{\mbox{\boldmath$\delta$\unboldmath}}
\newcommand \epsilonbf{\mbox{\boldmath$\epsilon$\unboldmath}}
\newcommand \zetabf{\mbox{\boldmath$\zeta$\unboldmath}}
\newcommand \etabf{\mbox{\boldmath$\eta$\unboldmath}}
\newcommand \iotabf{\mbox{\boldmath$\iota$\unboldmath}}
\newcommand \kappabf{\mbox{\boldmath$\kappa$\unboldmath}}
\newcommand \lambdabf{\mbox{\boldmath$\lambda$\unboldmath}}
\newcommand \mubf{\mbox{\boldmath$\mu$\unboldmath}}
\newcommand \nubf{\mbox{\boldmath$\nu$\unboldmath}}
\newcommand \xibf{\mbox{\boldmath$\xi$\unboldmath}}
\newcommand \pibf{\mbox{\boldmath$\pi$\unboldmath}}
\newcommand \rhobf{\mbox{\boldmath$\rho$\unboldmath}}
\newcommand \sigmabf{\mbox{\boldmath$\sigma$\unboldmath}}
\newcommand \taubf{\mbox{\boldmath$\tau$\unboldmath}}
\newcommand \upsilonbf{\mbox{\boldmath$\upsilon$\unboldmath}}
\newcommand \phibf{\mbox{\boldmath$\phi$\unboldmath}}
\newcommand \varphibf{\mbox{\boldmath$\varphi$\unboldmath}}
\newcommand \chibf{\mbox{\boldmath$\chi$\unboldmath}}
\newcommand \psibf{\mbox{\boldmath$\psi$\unboldmath}}
\newcommand \omegabf{\mbox{\boldmath$\omega$\unboldmath}}
\newcommand \Sigmabf{\hbox{$\bf \Sigma$}}
\newcommand \Upsilonbf{\hbox{$\bf \Upsilon$}}
\newcommand \Omegabf{\hbox{$\bf \Omega$}}
\newcommand \Deltabf{\hbox{$\bf \Delta$}}
\newcommand \Gammabf{\hbox{$\bf \Gamma$}}
\newcommand \Thetabf{\hbox{$\bf \Theta$}}
\newcommand \Lambdabf{\hbox{$\bf \Lambda$}}
\newcommand \Xibf{\hbox{\bf$\Xi$}}
\newcommand \Pibf{\hbox{\bf$\Pi$}}
\newcommand \abf{{\bf a}}
\newcommand \bbf{{\bf b}}
\newcommand \cbf{{\bf c}}
\newcommand \dbf{{\bf d}}
\newcommand \ebf{{\bf e}}
\newcommand \fbf{{\bf f}}
\newcommand \gbf{{\bf g}}
\newcommand \hbf{{\bf h}}
\newcommand \ibf{{\bf i}}
\newcommand \jbf{{\bf j}}
\newcommand \kbf{{\bf k}}
\newcommand \lbf{{\bf l}}
\newcommand \mbf{{\bf m}}
\newcommand \nbf{{\bf n}}
\newcommand \obf{{\bf o}}
\newcommand \pbf{{\bf p}}
\newcommand \qbf{{\bf q}}
\newcommand \rbf{{\bf r}}
\newcommand \sbf{{\bf s}}
\newcommand \tbf{{\bf t}}
\newcommand \ubf{{\bf u}}
\newcommand \vbf{{\bf v}}
\newcommand \wbf{{\bf w}}
\newcommand \xbf{{\bf x}}
\newcommand \ybf{{\bf y}}
\newcommand \zbf{{\bf z}}
\newcommand \rbfa{{\bf r}}
\newcommand \xbfa{{\bf x}}
\newcommand \ybfa{{\bf y}}
\newcommand \Abf{{\bf A}}
\newcommand \Bbf{{\bf B}}
\newcommand \Cbf{{\bf C}}
\newcommand \Dbf{{\bf D}}
\newcommand \Ebf{{\bf E}}
\newcommand \Fbf{{\bf F}}
\newcommand \Gbf{{\bf G}}
\newcommand \Hbf{{\bf H}}
\newcommand \Ibf{{\bf I}}
\newcommand \Jbf{{\bf J}}
\newcommand \Kbf{{\bf K}}
\newcommand \Lbf{{\bf L}}
\newcommand \Mbf{{\bf M}}
\newcommand \Nbf{{\bf N}}
\newcommand \Obf{{\bf O}}
\newcommand \Pbf{{\bf P}}
\newcommand \Qbf{{\bf Q}}
\newcommand \Rbf{{\bf R}}
\newcommand \Sbf{{\bf S}}
\newcommand \Tbf{{\bf T}}
\newcommand \Ubf{{\bf U}}
\newcommand \Vbf{{\bf V}}
\newcommand \Wbf{{\bf W}}
\newcommand \Xbf{{\bf X}}
\newcommand \Ybf{{\bf Y}}
\newcommand \Zbf{{\bf Z}}
\newcommand \Omegabbf{{\bf \Omega}}
\newcommand \Rssbf{{\bf R_{ss}}}
\newcommand \Ryybf{{\bf R_{yy}}}
\newcommand \Cset{{\cal C}}
\newcommand \Rset{{\cal R}}
\newcommand \Zset{{\cal Z}}
\newcommand{\otheta}{\stackrel{\circ}{\theta}}
\newcommand{\defeq}{\stackrel{\bigtriangleup}{=}}
\newcommand{\eqqcolon}{:=}
\newcommand{\oabf}{{\bf \breve{a}}}
\newcommand{\odbf}{{\bf \breve{d}}}
\newcommand{\oDbf}{{\bf \breve{D}}}
\newcommand{\oAbf}{{\bf \breve{A}}}
\renewcommand \vec{{\mbox{vec}}}
\newcommand{\Acalbf}{\bf {\cal A}}
\newcommand{\calZbf}{\mbox{\boldmath $\cal Z$}}
\newcommand{\feop}{\hfill \rule{2mm}{2mm} \\}
\newtheorem{theorem}{Theorem}[section]

%-------My Definition-------
\newcommand{\Rnum}{{\mathbb R}}
\newcommand{\Cnum}{{\mathbb C}}
\newcommand{\Znum}{{\mathbb Z}}
\newcommand{\Enum}{{\mathbb E}}
\newcommand{\Nnum}{{\mathbb N}}

\newcommand{\Ccal}{{\cal C}}
\newcommand{\Dcal}{{\cal D}}
\newcommand{\Hcal}{{\cal H}}
\newcommand{\Ocal}{{\cal O}}
\newcommand{\Rcal}{{\cal R}}
\newcommand{\Zcal}{{\cal Z}}
\newcommand{\Xcal}{{\cal X}}
\newcommand{\Ncal}{{\cal N}}
\newcommand{\Kcal}{{\cal K}}
\newcommand{\lcal}{{\cal l}}
\newcommand{\Scal}{{\cal S}}
\newcommand{\zzbf}{{\bf 0}}
\newcommand{\zebf}{{\bf 0}}

\newcommand{\eop}{\hfill $\Box$}

%---
\newcommand{\gss}{\mathop{}\limits}
\newcommand{\gs}{\mathop{\gss_<^>}\limits}
%---usage $$\gs_{H_1}^{H_0}$

\newcommand{\circlambda}{\mbox{$\Lambda$
             \kern-.85em\raise1.5ex
             \hbox{$\scriptstyle{\circ}$}}\,}

\newcommand{\tr}{\mathop{\rm tr}}
\newcommand{\var}{\mathop{\rm var}}
\newcommand{\cov}{\mathop{\rm cov}}
\newcommand{\diag}{\mathop{\rm diag}}
\def\rank{\mathop{\rm rank}\nolimits}
\newcommand{\ra}{\rightarrow}
\def\Pr{\mathop{\rm Pr}}
\def\Re{\mathop{\rm Re}}
\def\Im{\mathop{\rm Im}}

\def\submbox#1{_{\mbox{\footnotesize #1}}}
\def\supmbox#1{^{\mbox{\footnotesize #1}}}

%%%%%%%%   ``Theorem-like'' environments (defs, lemmas numbered like Theorems)
%
\newtheorem{Theorem}{Theorem}[section]
\newtheorem{Definition}[Theorem]{Definition}
\newtheorem{Proposition}[Theorem]{Proposition}
\newtheorem{Lemma}[Theorem]{Lemma}
\newtheorem{Corollary}[Theorem]{Corollary}
\newtheorem{Conjecture}[Theorem]{Conjecture}
%
% to label, reference them
%
\newcommand{\ThmRef}[1]{\ref{thm:#1}}
\newcommand{\ThmLabel}[1]{\label{thm:#1}}
\newcommand{\DefRef}[1]{\ref{def:#1}}
\newcommand{\DefLabel}[1]{\label{def:#1}}
\newcommand{\PropRef}[1]{\ref{prop:#1}}
\newcommand{\PropLabel}[1]{\label{prop:#1}}
\newcommand{\LemRef}[1]{\ref{lem:#1}}
\newcommand{\LemLabel}[1]{\label{lem:#1}}
%
%%%%%%%%

%\newcommand \abs{{\boldsymbol a}}
\newcommand \bbs{{\boldsymbol b}}
\newcommand \cbs{{\boldsymbol c}}
\newcommand \dbs{{\boldsymbol d}}
\newcommand \ebs{{\boldsymbol e}}
\newcommand \fbs{{\boldsymbol f}}
\newcommand \gbs{{\boldsymbol g}}
\newcommand \hbs{{\boldsymbol h}}
\newcommand \ibs{{\boldsymbol i}}
\newcommand \jbs{{\boldsymbol j}}
\newcommand \kbs{{\boldsymbol k}}
\newcommand \lbs{{\boldsymbol l}}
\newcommand \mbs{{\boldsymbol m}}
\newcommand \nbs{{\boldsymbol n}}
\newcommand \obs{{\boldsymbol o}}
\newcommand \pbs{{\boldsymbol p}}
\newcommand \qbs{{\boldsymbol q}}
\newcommand \rbs{{\boldsymbol r}}
\newcommand \sbs{{\boldsymbol s}}
\newcommand \tbs{{\boldsymbol t}}
\newcommand \ubs{{\boldsymbol u}}
\newcommand \vbs{{\boldsymbol v}}
\newcommand \wbs{{\boldsymbol w}}
\newcommand \xbs{{\boldsymbol x}}
\newcommand \ybs{{\boldsymbol y}}
\newcommand \zbs{{\boldsymbol z}}

\newcommand \Bbs{{\boldsymbol B}}
\newcommand \Cbs{{\boldsymbol C}}
\newcommand \Dbs{{\boldsymbol D}}
\newcommand \Ebs{{\boldsymbol E}}
\newcommand \Fbs{{\boldsymbol F}}
\newcommand \Gbs{{\boldsymbol G}}
\newcommand \Hbs{{\boldsymbol H}}
\newcommand \Ibs{{\boldsymbol I}}
\newcommand \Jbs{{\boldsymbol J}}
\newcommand \Kbs{{\boldsymbol K}}
\newcommand \Lbs{{\boldsymbol L}}
\newcommand \Mbs{{\boldsymbol M}}
\newcommand \Nbs{{\boldsymbol N}}
\newcommand \Obs{{\boldsymbol O}}
\newcommand \Pbs{{\boldsymbol P}}
\newcommand \Qbs{{\boldsymbol Q}}
\newcommand \Rbs{{\boldsymbol R}}
\newcommand \Sbs{{\boldsymbol S}}
\newcommand \Tbs{{\boldsymbol T}}
\newcommand \Ubs{{\boldsymbol U}}
\newcommand \Vbs{{\boldsymbol V}}
\newcommand \Wbs{{\boldsymbol W}}
\newcommand \Xbs{{\boldsymbol X}}
\newcommand \Ybs{{\boldsymbol Y}}
\newcommand \Zbs{{\boldsymbol Z}}

\newcommand \Absolute[1]{\left\lvert #1 \right\rvert}

% \title{Asynchronus Federated Learning Integrating Global Information and Local Information\\
\title{FedGSM: Efficient Federated Learning for LEO
Constellations with Gradient Staleness Mitigation\\
\thanks{Lingling Wu and Jingjing Zhang are with the Department of Communication Science and Engineering, Fudan University, Shanghai 200433, China (e-mail: 21210720245@m.fudan.edu.cn; jingjingzhang@fudan.edu.cn). This work has been supported by the National Natural Science Foundation of China Grant No. 62101134.}}

\author{\IEEEauthorblockN{Lingling Wu, Jingjing Zhang}}
% \date{March 2023}
\maketitle
\begin{abstract}
% Recent advances in space technology have equipped low Earth Orbit (LEO) satellites with highly sophisticated onboard computing capacity, enabling them to perform a range of functions and AI applications. Implementing Federated Learning (FL) on LEO satellites can facilitate the development of AI by enabling collaborative training of a global ML model without the need for sharing a large dataset. However, the intermittent connectivity between the satellites and ground stations can lead to stale gradients and unstable learning in asynchronous FL algorithms, limiting the learning performance. To address this challenge, we propose FedGSM, a novel asynchronous Federated Learning algorithm that introduces a compensation term to mitigate staleness. Our simulations show that FedGSM outperforms state-of-the-art algorithms for both IID and non-IID datasets, demonstrating its effectiveness and advantages.
Recent advancements in space technology have equipped low Earth Orbit (LEO) satellites with the capability to perform complex functions and run AI applications. Federated Learning (FL) on LEO satellites enables collaborative training of a global ML model without the need for sharing large datasets. However, intermittent connectivity between satellites and ground stations can lead to stale gradients and unstable learning, thereby limiting learning performance. In this paper, we propose FedGSM, a novel asynchronous FL algorithm that introduces a compensation mechanism to mitigate gradient staleness. FedGSM leverages the deterministic and time-varying topology of the orbits to offset the negative effects of staleness. Our simulation results demonstrate that FedGSM outperforms state-of-the-art algorithms for both IID and non-IID datasets, underscoring its effectiveness and advantages. We also investigate the effect of system parameters. 
\end{abstract}


\section{Introduction}

The advancement of satellite communication, particularly in low Earth orbit (LEO) satellites, has made it feasible to implement artificial intelligence (AI) in satellite communication scenarios for various purposes, such as anti-jamming \cite{han2020dynamic}, network traffic forecasting \cite{na2018distributed}, weather forecasting, and soil moisture monitoring \cite{GNSS}. However, downloading data and training machine learning (ML) models on the ground poses significant challenges, including communication and computation overhead \cite{javan2021review}, as well as privacy concerns about exposing client data \cite{coffer2020balancing}. Federated learning (FL) \cite{FedAvg} provides an effective solution to these challenges when implemented in satellite constellations. Each satellite trains a local ML model without uploading its data. Instead, they send the trained model parameters to a parameter server (PS), which can be a ground station (GS). The PS then combines them into a global model and sends it back to all the satellites. This process repeats until the global model converges.

 % By following this approach, FL ensures that data privacy is maintained and that the communication overhead is minimized, making it an attractive option for training ML models on LEO satellites

% In this approach, each satellite obtains an initial model from the GSand trains the model independently, without exposing local data. The local model is then sent to the GSto update the global model, which initializes the next local training.

\textbf{Challenges.}
% However, FL on satellites faces two significant challenges. First, the connectivity between the satellites and the GSis highly intermittent. A typical LEO satellite has a visibility period of 5-20 minutes \cite{perez1998introduction} and an orbit period of 90-120 minutes \cite{2000km20min}, meaning the communicable period is brief and only a small fraction of the orbital period. Second, satellites at lower altitudes revisit the GSmore frequently, and vice versa. As shown below, the performance of FL algorithms would be hurt, regardless of whether synchronous or asynchronous algorithms are used. 
Implementing FL on satellites, however, poses two significant challenges. Firstly, the connectivity between the satellites and the GS is highly intermittent due to the typical LEO satellite's visibility period of 5-20 minutes and orbit period of 90-120 minutes \cite{perez1998introduction,2000km20min}. Secondly, the performance of FL algorithms is impacted by the varying frequency of revisits by satellites at different altitudes to the ground station, regardless of whether synchronous or asynchronous algorithms are utilized.

% However, implementing FL on satellites faces two significant challenges. The first challenge is the highly intermittent connectivity between the satellites and the ground station. For example, a typical LEO satellite has a visibility period of only 8-11 minutes \cite{abrishamkar1996pcs} and an orbit period of 90-110 minutes \cite{cakaj2021parameters}, which means that the communicable period is only a small fraction of the orbital period. The second challenge is that satellites at lower altitudes revisit the GSmore frequently, while the opposite is true for higher altitudes. Regardless of whether synchronous or asynchronous algorithms are used, these factors can significantly harm the performance of FL algorithms, as shown below.

In synchronous algorithms such as FedAvg \cite{FedAvg}, the global model is sent to all satellites by the ground station, which then waits for the local models to be returned, resulting in significant time wastage, particularly when there are stragglers with low re-entry frequencies. Asynchronous algorithms such as FedBuff \cite{fedbuff} involve sending a stale global model to a satellite when it enters a visible period and collecting only a few local models for the global update. Although this approach may speed up the convergence of the learning process, the gradient staleness of the initial model in local training \cite{dutta2018slow} can limit performance significantly. Additionally, the learning process may not be robust due to incomplete client participation \cite{zhou2022towards}.

\textbf{Related work.}
% For accelerating the process of broadcasting and retrieving models in synchronous algorithms, \cite{on-board} proposed to exchange models by inter-satellite communication, and \cite{fedhap} introduced high-altitude platforms (HAPs). However, these approaches require additional costs for equipment deployment and result in increased inter-satellite or HAP-ground communication overheads. For alleviating the staleness in asynchronous algorithms, \cite{fedsatschedule} proposes to complete the learning task and the communication task in a visible period. However, this is impractical since the visible period is short and valuable for communication if there are multiple satellites to communicate simultaneously, not to mention for training and communicating simultaneously. Besides, \cite{fedspace} designs an adaptive aggregation scheduler to make a trade off between synchronous algorithm and asynchronous algorithm. But it requires for some data collected in the ground station, which is against the principle of privacy.
Multiple approaches have been proposed to address the challenges faced by FL on satellites. To accelerate the process of broadcasting and retrieving models in synchronous algorithms, inter-satellite communication has been used in \cite{on-board}, and high-altitude platforms (HAPs) have been introduced in \cite{fedhap}. However, these approaches require additional costs for equipment deployment and result in increased inter-satellite or HAP-ground communication overheads. To alleviate the staleness in asynchronous algorithms, \cite{fedsatschedule} proposes to complete the learning task and the communication task in a visible period. However, this approach is impractical since the visible period is short and valuable for communication when multiple satellites need to communicate simultaneously. Moreover, \cite{fedspace} designs an adaptive aggregation scheduler that balances the trade-off between synchronous and asynchronous algorithms. However, this approach requires some data collected in the ground station, which violates the principle of privacy.

\textbf{Contributions.}
% To handle the problems of the staleness of information and the instability of learning process, we propose FedGSM, an \emph{asynchronous}  \emph{Federated} learning algorithm integrating \emph{global} information and \emph{local} information. The global information of model difference plays a role like momentum, alleviating the instability caused by asynchronous aggregation and accelerate the learning process. The local information of gradient difference plays a role of second order information like hessian information to compensate for the staleness of local model. Simulation results show that compared with the baselines, FedGSM improves the accuracy by $5.93\%$ and $7.10\%$ for IID and non-IID CIFA-10 dataset, respectively.
% We present FedGSM, an asynchronous Federated Learning algorithm that addresses the issues of gradient staleness and learning instability. FedGSM leverages historic local-information to mitigate the instability induced by asynchronous aggregation. It uses the difference of the local model difference, which offsets to staleness effect, to compensate for the staleness of local models. Our simulation results demonstrate that FedGSM significantly enhances accuracy improvement on both the IID and non-IID CIFA-10 datasets, respectively, as compared to the baseline algorithms.
% information as momentum to alleviate instability while local gradient difference information acts as second-order information, akin to hessian information, to compensate for the staleness of local models. 
We propose FedGSM, an asynchronous Federated Learning algorithm that effectively addresses the issues of gradient staleness and learning instability. It utilizes the difference between consecutive local models to correct the local updates, thereby mitigating the negative effects of gradient staleness. Our simulation results demonstrate that FedGSM outperforms state-of-the-art algorithms on both IID and non-IID CIFAR-10 datasets, leading to significant accuracy improvements.

The remainder of this paper is organized as follows. In Section \ref{sec2}, we present the communication model and the framework of the asynchronous FL algorithm. We then introduce the proposed FedGSM algorithm in Section \ref{sec:FedGSM}. Simulation results are presented in Section \ref{sec:sim}, and finally, Section \ref{sec:con} concludes the paper.

\section{System Model}\label{sec2}

This section begins with an introduction to the satellite-to-ground communication model, followed by an overview of the general framework for the asynchronous FL algorithm.

\subsection{Satellite Communication model for FL} \label{sec:com_model}

% We investigate a satellite-ground asynchronous FL cooperation paradigm in which the $N$ satellites in $m$ orbital planes function as the clients and a single GSserves as the server, as shown in Fig. \ref{fig:SA-GS}. Note that both the GSand satellite constellations can be altered.

\begin{figure}[t!]
\centering
\setlength{\abovecaptionskip}{-1.85cm}
% \includesvg[inkscapelatex=false,width=3in]{./Result_Plot/SA-GS_v7.svg}
% \hspace{-2cm}%左移
% \includesvg[inkscapelatex=false,width=3.5in]{./Result_Plot/SA-GS_v8.svg}
% \includesvg[inkscapelatex=false,width=3.5in]{./SA-GS_0412.svg}
\includegraphics[width=3.5in,keepaspectratio=false]{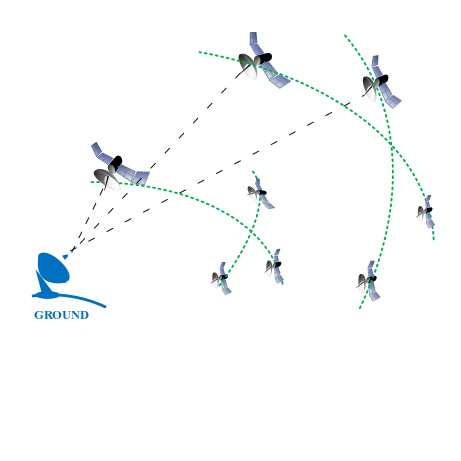}
\caption{A uniform constellation of $N=8$ satellites in $m=4$ orbits at two different altitudes.}
\label{fig:SA-GS}
\end{figure}

As depicted in Fig. \ref{fig:SA-GS}, we examine a satellite-ground communication model consisting of a constellation of $N$ satellites from the set $\Ncal=\{1,2,\cdots,N\}$, operating in $m$ orbital planes, and a single ground station (GS). The constellation can be configured in various ways, such as the Walker constellation \cite{walker1984satellite}, the SpaceX constellation \cite{del2019technical}, and so on.

% In an earth-centered inertial coordinate system, each satellite $k \in \Ncal =\{1, 2, \cdots, N\}$ has trajectory $\rbf_{n}\left(t\right) = \left[r_n^x(t), r_n^y(t), r_n^z(t)\right]$ and the GShas a trajectory $\rbf_{g}\left(t
% \right) = \left[r_g^x(t), r_g^y(t), r_g^z(t)\right]$, where $t$ is real-world time. Satellite $k$ could communicate to the GSonly when it is visible to the ground station, i.e., 
% \ben
% \alpha_{k,g} = \angle\left(\rbf_{g}(t),\rbf_{n}(t) - \rbf_{g}(t)\right) \leq \frac{\pi}{2} - \alpha_{min},
% \een
% where $\alpha_{min}$ is the minimum elevation angle determined by the link budget requirement of the system. Note that the trajectories $\rbf_{n}\left(t\right)$ and $\rbf_{g}\left(t\right)$ are time-varying due to the revolution of satellites and the rotation of Earth, respectively. Consequently, the satellite may be invisible to the GSwhen $\alpha_{k,g} > \frac{\pi}{2} - \alpha_{min}$. 

In an earth-centered inertial coordinate system, each satellite $n \in \Ncal$ has a 3D trajectory $\rbf_{n}\left(t\right)$ and the GS has a 3D trajectory $\rbf_{g}\left(t\right)$, where $t$ is real-world time. Each satellite $n$ can only communicate with the GS when it is visible to the GS. This occurs when the angle between the line of sight from the GS to the satellite and the local vertical direction at the GS is greater than or equal to $\alpha_{min}$, i.e., we have 
% \ben
% \alpha_{n,g} = \angle\left(\rbf_{g}(t),\rbf_{n}(t) - \rbf_{g}(t)\right) \leq \frac{\pi}{2} - \alpha_{min},
% \een
\begin{equation}
\setlength{\abovedisplayskip}{3pt}
\setlength{\belowdisplayskip}{3pt}
\alpha_{n,g} = \angle\left(\rbf_{g}(t),\rbf_{n}(t) - \rbf_{g}(t)\right) \leq \frac{\pi}{2} - \alpha_{min},
\end{equation}
where $\alpha_{min}$ is the minimum elevation angle required by the link budget of the system. 
% See \cite{ali1999predicting} for more details about the prediction of the visible period.
% Note that the trajectories $\rbf_{n}\left(t\right)$ and $\rbf_{g}\left(t\right)$ are time-varying due to the revolution of satellites and the rotation of Earth, respectively. Consequently, a satellite may be invisible to the GSwhen $\alpha_{n,g} > \frac{\pi}{2} - \alpha_{min}$. The visible pattern is predictable, see \cite{ali1999predicting} for more details.

\begin{figure}[ht]
\centering
% \setlength{\abovecaptionskip}{-6cm}
% \includesvg[inkscapelatex=false,width=3.3in]{./plot_period_v6.svg}
\includegraphics[keepaspectratio=false,width=3.3in]{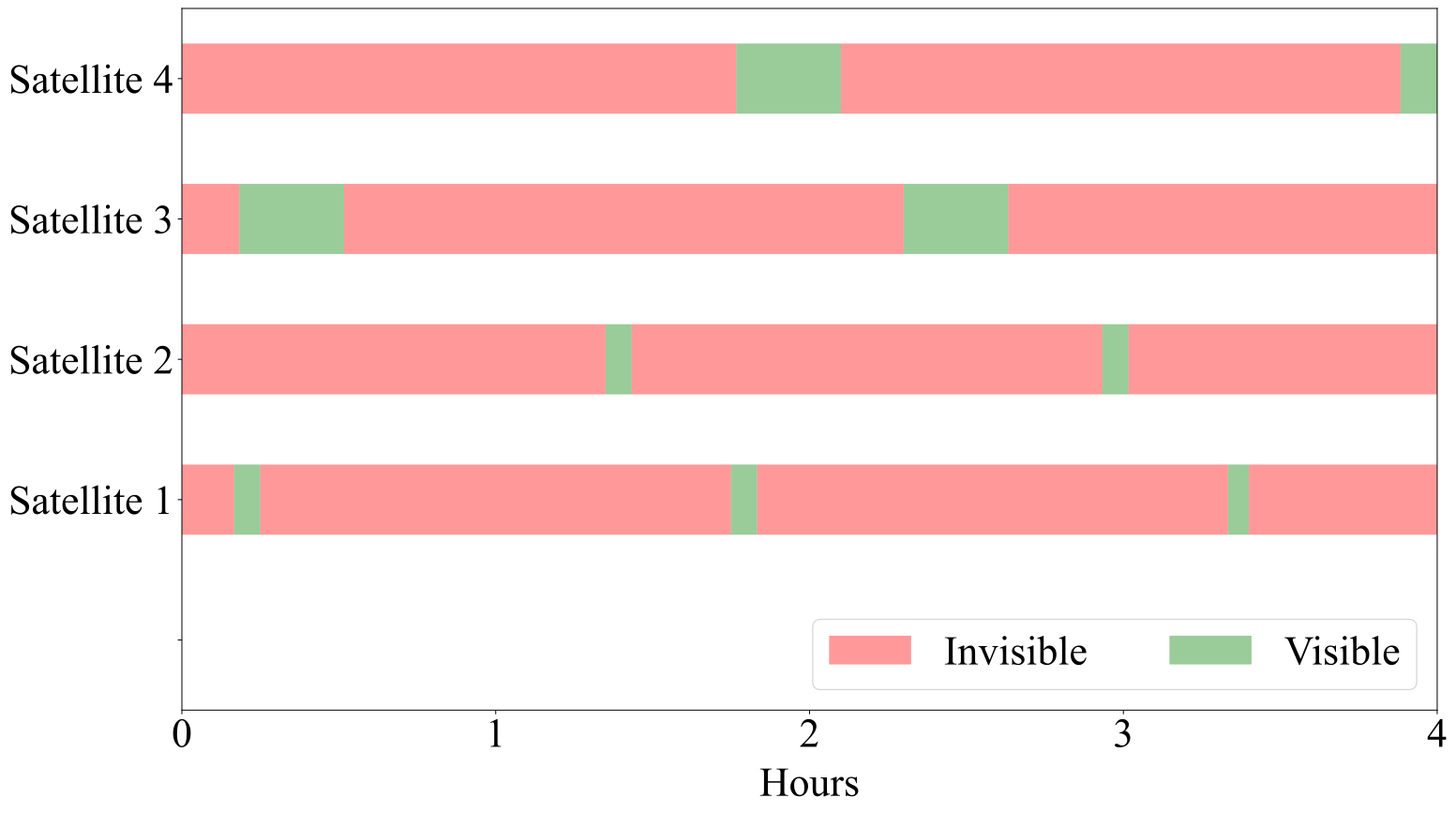}
\caption{Visible patterns of satellites. Satellites 1 and 2 are at an altitude of $500$ Km, Satellites 3 and 4 are at an altitude of $2000$ Km, and the GS is at the North Pole.}
\label{fig:SA_period}
\end{figure}

The visible period of a LEO satellite relative to a GS is usually short compared to its orbital period, as illustrated in Fig. \ref{fig:SA_period}. This means that the satellite-ground communication is intermittent and brief, leading to communication staleness that can be detrimental to training convergence. Moreover, higher-altitude satellites revisit the GS less frequently, resulting in longer idle periods and lower convergence rates in synchronous FL. Therefore, asynchronous FL is a more suitable approach for satellite-ground communication scenarios.

\subsection{Asynchronous FL Algorithm for Satellite Communication} \label{sec:AFL}

We utilize the satellite-ground communication model to carry out FL tasks. Specifically, the $N$ satellites operate as clients with their respective local datasets, while the single GS functions as the server. Our objective is to optimize the following problem
\bea \label{eq:obj}
\min_{\Wbf\in \Rnum^{d}}\ f(\Wbf) 
&\eqqcolon \frac{1}{N} \sum_{n=1}^{N} p_n F_n\left(\Wbf\right)\\
&= \frac{1}{N} \sum_{n=1}^{N}\sum_{\xbf \in \Dcal_{n}} \frac{p_n}{d_n}{l_n\left(\Wbf;\xbf \right)},
\eea
% where $\Wbf$ is the model parameter vector, $N$ is the number of satellites, $F_n\left(\Wbf\right)$ measures the average loss of $\Wbf$ on the $n$-th satellite's local dataset $\Dcal_{n}$ with the cardinality denoted by $d_n$, and $p_n = d_n/(\sum_{n=1}^{N} d_n)$ weights the importance of $\Dcal_{n}$. To mitigate the effect of staleness, we apply an asynchronous stochastic gradient descent (SGD) mechanism with $K \leq N$ sequential satellites participating in the aggregation at each round after multiple local updates.
where $\Wbf$ is the model parameter vector; $N$ is the number of satellites; $F_n(\Wbf)$ measures the average loss of $\Wbf$ on the $n$-th satellite's local dataset $\Dcal_{n}$, which has cardinality $d_n$; and $p_n = d_n/(\sum_{n=1}^{N} d_n)$ weights the importance of $\Dcal_{n}$. To mitigate the negative effects of communication staleness, we apply an asynchronous stochastic gradient descent (SGD) mechanism, in which $K \leq N$ sequential satellites participate in the aggregation at each round after multiple local updates.

\begin{figure}[t!]
\centering
\setlength{\abovecaptionskip}{-1.5cm}
% \hspace{-2cm}%左移
% \includesvg[inkscapelatex=false,width=3.5in]{./Result_Plot/AFL.svg}
% \includesvg[inkscapelatex=false,width=3.5in]{./Result_Plot/AFL_2.svg}
% \includesvg[inkscapelatex=false,width=3.5in]{AFL_3.svg}
\includegraphics[width=3.5in,keepaspectratio=false]{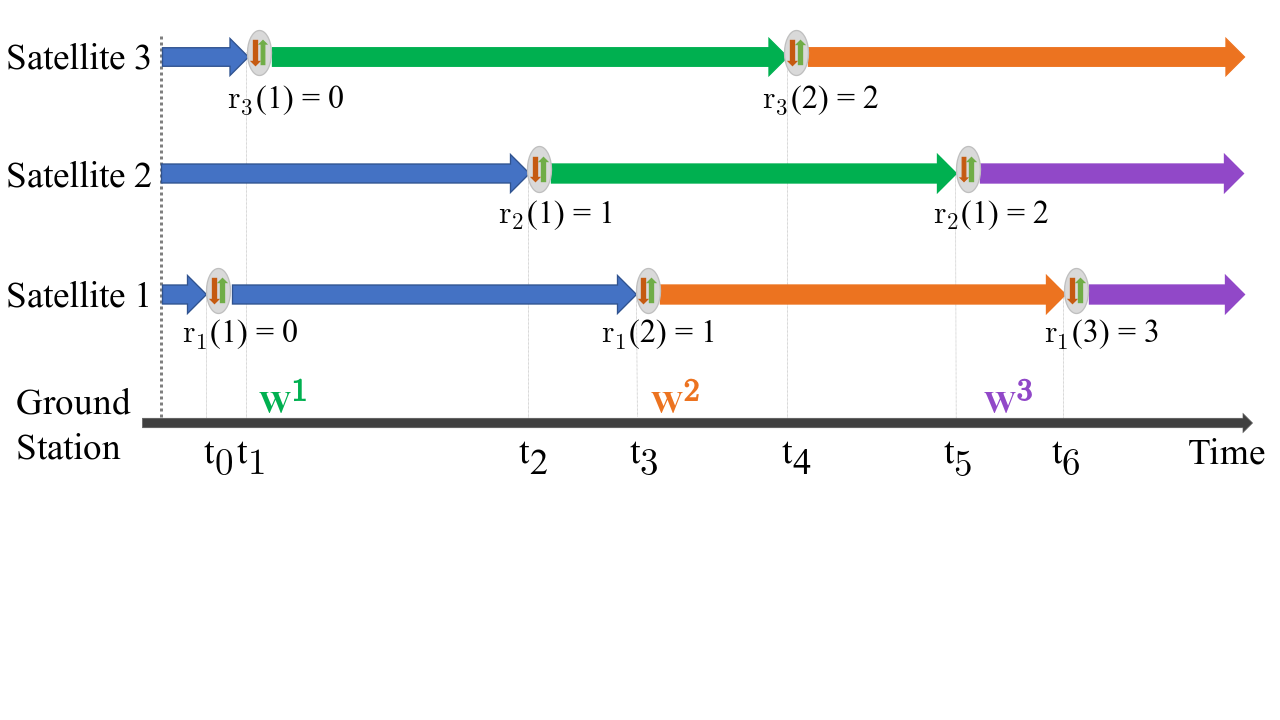}
\caption{The timing diagram of the asynchronous FL in an illustrative example with $N = 3, K = 2$.}
\label{fig:AFL}
\end{figure}

To elaborate on the training process, we refer to the $i$-th round as the time slot $[t_{Ki}, t_{K(i+1)})$, where $t_{Ki}$ is the instant when the $Ki$-th satellite becomes visible. Each satellite $n\in \Ncal$ would record its participation in the training process by maintaining the vector ${\rbf}_n$. For example, if round $i$ is the $j$-th participation for satellite $n$, i.e., we have $n\in \Scal^i$, then the $j$-th element of the vector ${\rbf}_n$ is $r_n\left(j\right)=i$. Fig. \ref{fig:AFL} illustrates the time diagram of the asynchronous FL algorithm in an example with $N=3$ and $K=2$.

In each round $i$, the GS waits to receive the local updates $\{{\bf \bm{\nabla}}_n^i, n \in \Scal^i\}$ from $K$ satellites in the set $\Scal^i$ with a cardinality of $K$. The updates arrive in time order, and it is important to note that each update ${\bf \bm{\nabla}}_n^i$ is a function of the computed gradients. For the first $K-1$ satellites in the set $\Scal^i$, the GS receives each update ${\bf \bm{\nabla}}_n^i$ and sends back the current up-to-date global model $\Wbf^i$ to them. Once it receives ${\bf \bm{\nabla}}_n^i$ from the $K$-th satellite, it aggregates and updates the global model as follows:
\ben \label{eq:FL}
\Wbf^{i \text{+} 1} = \Wbf^{i} + \eta_{g}\sum_{n\in \Scal^i} \frac{p_n}{\sum_{n\in \Scal^i} p_n}{\bf \bm{\nabla}}_n^i,
\een
where $\eta_{g}$ is the global learning rate. Then, the updated model $\Wbf^{i \text{+} 1}$ would be sent to the $K$-th satellite.

Once visible, each satellite $n\in \Scal^i$ sends ${\bf \bm{\nabla}}_n^i$ first and also receives the up-to-date global model. More precisely, 
if $n$ is the $K$-th element of the set $\Scal^i$, it would receive the updated global model $\Wbf^{i+1}$ (see also Eq.~\ref{eq:FL}) and set the initial local model for the next participating round as $\Wbf_n^{r_n\left(j\right),0} = \Wbf^{i+ 1}$. For other satellites in the set $\Scal^i$, we have $\Wbf_n^{r_n\left(j\right),0} = \Wbf^{i}$ in the sense that the global model has not been updated yet. After communication with the ground station, each satellite begins local SGD for $E$ epochs, given by:
% \ben \label{eq:local_update}
% \Wbf_{n}^{r_n\left(j\right), e} = \Wbf_{n}^{r_n\left(j\right), e\shortminus 1} - \eta_{l}g_{n}\left(\Wbf_{n}^{r_n\left(j\right), e\shortminus 1}\right),
% \een
\ben \label{eq:local_update}
\Wbf_{n}^{r_n\left(j\right), e} = \Wbf_{n}^{r_n\left(j\right), e\shortminus 1} - \eta_{l}g_{n}^{r_n\left(j\right), e\shortminus 1},
\een
for $e = 1, 2, \cdots, E$, where $\eta_l$ is the local learning rate, and $g_{n}^{r_n\left(j\right), e\shortminus 1} = \nabla F_{n}\left(\Wbf_{n}^{r_n\left(j\right), e\shortminus 1}\right)$ is the computed local gradient of each satellite $n$ at the $e$-th epoch. The training processing continues until the desired convergence criterion is satisfied. 
% for $e = 1, 2, \cdots, E$, where $\eta_l$ is the local learning rate, and $g_{n}\left(\Wbf_{n}^{r_n\left(j\right), e\shortminus 1}\right)$ is the computed local gradient of each satellite $n$ at the $e$-th epoch. The training processing continues until the desired convergence criterion is satisfied. 

% Based on the training results, each satellite $n$ would send the local update ${\bf\bm{\nabla}}_n^{r_n\left(j\text{+}1\right)}$ as a function of ${\Wbf}_{n}^{r_n\left(j\right),e}$ to the GS in its next visible period, corresponding to the $r_n\left(j\text{+}1\right)$-th round. For example, in the implementation of FedBuff \cite{fedbuff} on satellites, ${\bf\bm{\nabla}}_n^{r_n(j\text{+}1)}$ is indeed the so-called pseudo-gradient:
% \ben \label{eq:pse_grad}
% \gbf_n^{r_n\left(j\right)} \triangleq \Wbf_n^{r_n\left(j\right), E} - \Wbf_n^{r_n\left(j\right), 0}.
% \een

% Note that for each satellite $n$, the GShas updated $r_n\left(j\text{+}1\right) - r_n\left(j\right)$ times before its next visible period. In other words, the information returned by each satellite $n$ is computed at a stale value of the global parameter. The staleness $\tau_{n}^{r_n\left(j\text{+}1\right)}$ hence can be given as 
% \ben
% \tau_{n}^{r_n\left(j\text{+}1\right)} \triangleq r_n\left(j\text{+}1\right) - r_n\left(j\right).
% \een
Note that for each satellite $n$, the GS has updated $r_n\left(j\right) - r_n\left(j\shortminus1\right)$ times before its visible period. In other words, the information returned by each satellite $n$ is computed at a stale value of the global parameter. The staleness $\tau_{n}^{r_n\left(j\right)}$ hence can be given as 
\ben
\tau_{n}^{r_n\left(j\right)} \triangleq r_n\left(j\right) - r_n\left(j\shortminus1\right).
\een
For an asynchronous FL, the staleness $\tau_{n}^{r_n\left(j\right)}$ can be high because of the periodical and lengthy invisible period. Furthermore, higher-altitude satellites revisit the GS less frequently, resulting in a higher staleness. Staleness $\tau_{n}^{r_n\left(j\right)}$ can hence decelerate convergence and even intensify divergence of the global model significantly.

% \begin{remark}
% Staleness $\tau_{n}^{r_n\left(j\right)}$ can decelerate convergence and even intensify divergence of the global model. Additionally, asynchronous algorithms can be unstable due to data heterogeneity and staleness. Specifically, the historic global consensus regarding the descending direction can be destroyed by incomplete local information, as only stale information from $K \leq N$ clients is used per round.
% \end{remark}

\section{FedGSM} \label{sec:FedGSM}

In this section, we introduce a novel asynchronous algorithm with gradient staleness mitigation named FedGSM. Given that the topology of the satellite model is predictable, the staleness of each satellite is periodic and stable. FedGSM exploits this characteristic to offset the negative effects of staleness by leveraging the difference between the computed local results in two consecutive rounds. Additionally, all the required information can be easily computed on the client side and transmitted back to the server without incurring any additional communication overhead.

% The main idea behind FedGSM is to address the issue of the staled local information by incorporating historic gradient differences and global increment information, thereby preventing divergence from the global consensus and accelerating convergence in a stable manner. Importantly, all the necessary information can be conveniently computed on the client side and sent back to the server without incurring additional communication overhead.

\begin{algorithm}[t!]
\caption{\bf FedGSM-Server (GSOperation)}\label{alg:FedGSM-SERVER}
\algorithmicrequire{ model $\Wbf^{0}$, client number $N$, data importance $\pbf = [p_1, \cdots, p_N]$, global learning rate $\eta_g$, buffer size $K$.}

\algorithmicInit{${\bf \bm{\nabla}}^{i} = {\bf 0}$, $k = 0$, $p = 0$, $i = 0$.}

\algorithmicensure{ FL-trained global model $\Wbf$.}
\begin{algorithmic}[1]%每行显示行号  
\Repeat
\State Wait for any satellite. Upon connection to satellite $n$:
\If {received update ${\bf \bm{\nabla}}_n^{i}$ from satellite $n$}
\State ${\bf \bm{\nabla}}^{i} = {\bf \bm{\nabla}}^{i} + p_n{\bf \bm{\nabla}}_n^{i}$;
\State $p = p + p_n$, $k = k + 1$;
\If {$k == K$}
%\State ${\bf \bm{\nabla}} = \frac{1}{p}{\bf \bm{\nabla}}$
% \State $\Wbf^{i+1} = \Wbf^{i} - \frac{1}{p}\eta_g{\bf \bm{\nabla}}^{i}$;
\State $\Wbf^{i+1} = \Wbf^{i} + \frac{1}{p}\eta_g{\bf \bm{\nabla}}^{i}$;
\State $\bm{\nabla}^{i} = {\bf 0}$, $p = 0$, $k = 0$, $i = i+1$;
\EndIf
\State send global model $\Wbf^{i}$ to the satellite $n$.
\EndIf
\State Terminate connection to satellite $n$.
\State Executes FedGSM-Client on satellit $n$.
\Until {Convergence}
\end{algorithmic}    
\end{algorithm}

\begin{algorithm}[t!]
\caption{\bf FedGSM-Client (Satellites Operation)}\label{alg:FedGSM-Client}
\algorithmicrequire{ global model $\Wbf$, local learning rate $\eta_l$, local epoch $E$}.

\algorithmicInit{$\Wbf_n^{r_n\left(j\shortminus1\right),E} = \bf{0}$, $e = 0$, ${\rbf}_n = \bf{0}$ }

\algorithmicensure{ client update ${\bf \bm{\nabla}}_n^{r_n\left(j\text{+}1\right)}$}

\begin{algorithmic}[1]%每行显示行号  
\State $r_n\left(j\right) = i$;
\If{received the global model ${\Wbf}^{i}$ from the GS}
% \State $r_n\left(j\right) = i$;
\State $\Wbf_n^{r_n\left(j\right),0} = {\Wbf}^{i}$;
\For {$e = 1:E$}
\State $\Wbf_n^{r_n\left(j\right),e} = \Wbf_n^{r_n\left(j\right),e\shortminus 1} - \eta_l g_n^{(r_n\left(j\right),e\shortminus 1)}$;
\EndFor
% \State $\gbf_n^{r_n\left(j\right)} = {\Wbf}_n^{r_n\left(j\right),E} - \Wbf^{r_n\left(j\right),0}$
% \State ${\bf \bm{\nabla}}_n^{r_n\left(j\right)} = \gbf_n^{r_n\left(j\right)} + \Wbf_n^{r_n\left(j\right),E} - \Wbf_n^{r_n\left(j\shortminus1\right),E}$;
% \State ${\bf \bm{\nabla}}_n^{i} = {\Wbf}^{i} + \Wbf_n^{old} - 2\Wbf_n^{i,E}$;
\State ${\bf \bm{\nabla}}_n^{r_n\left(j\text{+}1\right)} =  2\Wbf_n^{r_n\left(j\right),E} - {\Wbf}^{r_n\left(j\right),0} - \Wbf_n^{r_n\left(j\shortminus1\right),E}$;
\State $\Wbf_n^{r_n\left(j\shortminus1\right),E} = \Wbf_n^{r_n\left(j\right),E}$;
\EndIf
% \If{${\bf \bm{\nabla}}_n^{i}$ is ready}
\State send ${\bf \bm{\nabla}}_n^{r_n\left(j\text{+}1\right)}$ to the ground station in the $r_n\left(j\text{+}1\right)$-th round.
% \EndIf
\end{algorithmic}    
\end{algorithm}
% \begin{algorithm}[t!]
% \caption{\bf FedGSM-Client (Satellites Operation)}\label{alg:FedGSM-Client}
% \algorithmicrequire{ global model $\Wbf$, local learning rate $\eta_l$, local epoch $E$}.

% \algorithmicInit{$\Wbf_n^{old} = 0$, $e = 0$, }

% \algorithmicensure{ client update ${\bf \bm{\nabla}}_n^{i}$}

% \begin{algorithmic}[1]%每行显示行号  
% \If{received the global model ${\Wbf}^{i}$ from the GS}
% % \State $r_n\left(j\right) = i$;
% \State $\Wbf_n^{i,0} = {\Wbf}^{i}$;
% \For {$e = 1:E$}
% \State $\Wbf_n^{i,e} = \Wbf_n^{i,e\shortminus 1} - \eta_l g(\Wbf_n^{i,e\shortminus 1})$;
% \EndFor
% \State $\gbf_n^{i} = {\Wbf}_n^{i,E} - \Wbf^i$
% \State ${\bf \bm{\nabla}}_n^{i} = \gbf_n^{i} + \Wbf_n^{i,E} - \Wbf_n^{old}$;
% % \State ${\bf \bm{\nabla}}_n^{i} = {\Wbf}^{i} + \Wbf_n^{old} - 2\Wbf_n^{i,E}$;
% \State $\Wbf_n^{old} = \Wbf_n^{i,E}$;
% \EndIf
% % \If{${\bf \bm{\nabla}}_n^{i}$ is ready}
% \State send ${\bf \bm{\nabla}}_n^{i}$ to the ground station.
% % \EndIf
% \end{algorithmic}    
% \end{algorithm}

Generally speaking, FedGSM follows the asynchronous FL framework described in Section \ref{sec:AFL}. More precisely, for each satellite $n \in \Scal^i$ that participates in round $i$, it sends the local update ${\bf \bm{\nabla}}_n^{r_n\left(j\right)}$ to the GS, given as 
% \bea \label{eq:up_dire1}
% {\bf \bm{\nabla}}_n^{r_n\left(j\right)} 
% &\triangleq \gbf_n^{r_n(j\shortminus 1)} + \Wbf_n^{r_n(j\shortminus 1), E} - \Wbf_n^{r_n(j\shortminus 2), E}.
% \eea
\bea \label{eq:up_dire1}
{\bf \bm{\nabla}}_n^{r_n\left(j\right)} 
&\triangleq 2\Wbf_n^{r_n(j\shortminus 1), E} - \Wbf_n^{r_n(j\shortminus 1), 0} - \Wbf_n^{r_n(j\shortminus 2), E}.
\eea
The GS sequentially communicates with $K$ satellites in the set $\Scal^i$. For the first $K-1$ satellites, the GS receives their updates ${\bf\bm{\nabla}}_n^i$ and returns the current global model $\Wbf^i$. Each participating satellite receives $\Wbf^i$ and sets its local model as $\Wbf_n^{r_n\left(j\right),0}=\Wbf^i$ before computing the next round. Once the GS receives the $K$-th satellite's local updates ${\bf\bm{\nabla}}_n^i$, it aggregates and updates the global model using Eq.~(\ref{eq:FL}) and sends back ${\Wbf}^{i+1}$ to the $K$-th satellite. This satellite then sets its local model as $\Wbf_n^{r_n\left(j\right),0}=\Wbf^{i+1}$ to continue computations. The next round then repeats the operation. 

 The server-side operations and the client-side operations are summarized in Algorithm \ref{alg:FedGSM-SERVER} and Algorithm \ref{alg:FedGSM-Client}, respectively.
% With the new global model received in $i$-th round as the initial model $\Wbf_{n}^{r_n\left(j\right),0}$, each satellite $n \in \Scal^i$ performs SGD for $E$ epochs and obtains new local updates 
% in the method like (\ref{eq:local_update}). Note that $j$ is the index satisfying $r_n(j) = i$ in the participation round sequence $\bf{r}_n$ of satellite $n$. 

% In particular, we define the local update ${\bf \bm{\nabla}}_n^{r_n\left(j\right)}$ of satellite $n$ sent to the GS in $r_n\left(j\right)$-th round as follows:
% \bea \label{eq:up_dire1}
% {\bf \bm{\nabla}}_n^{r_n\left(j\right)} 
% &\triangleq \gbf_n^{r_n(j\shortminus 1)} + \Wbf_n^{r_n(j\shortminus 1), E} - \Wbf_n^{r_n(j\shortminus 2), E}.
% \eea
% Clearly the only cost is to store $\Wbf_n^{r_n(j\shortminus 2), E}$ at satellite $n$.
% $\gbf_n^{r_n(j\shortminus 1)} = {\Wbf}_n^{r_n\left(j\shortminus1\right),E} - {\Wbf}_n^{r_n\left(j\shortminus1\right),0} $ is the pseudo-gradient, which trained during the time after terminating connection in $r_n\left(j\shortminus1\right)$-th round and before the start of the connection in $r_n\left(j\right)$-th.
% Besides, (\ref{eq:up_dire1}) introduced a compensation term. For simplicity, we note the compensation term ${\bf\Delta}_n^{r_n\left(j\right)}$, i.e.
% \bea \label{compe_term}
% {\bf\Delta}_n^{r_n\left(j\right)} 
% &= \Wbf_n^{r_n(j\shortminus 1), E} - \Wbf_n^{r_n(j\shortminus 2), E}\\
% &= {\Wbf}_n^{r_n\left(j\shortminus 1\right), 0} - {\Wbf}_n^{r_n\left(j\shortminus 2\right), 0} + \gbf_n^{r_n\left(j\shortminus 1\right)}-\gbf_n^{r_n\left(j\shortminus 2
% \right)}.
% \eea

\begin{remark}
The local update ${\bf \bm{\nabla}}_n^{r_n\left(j\right)}$ of each satellite $n$ sent to the GS can be rewritten as follows:
\bea \label{eq:up_dir}
{\bf \bm{\nabla}}_n^{r_n\left(j\right)} 
&\triangleq \Wbf_n^{r_n(j\shortminus 1), E} - \Wbf_n^{r_n(j\shortminus 1), 0}+ 
{\bf \Delta}_n^{r_n\left(j\right)},
\eea
with ${\bf\Delta}_n^{r_n\left(j\right)}=\Wbf_n^{r_n(j\shortminus 1), E}- \Wbf_n^{r_n(j\shortminus 2), E}$. 
% 
% Instead of directly sending the incremental $\Wbf_n^{r_n(j-1), E} - \Wbf_n^{r_n(j-1), 0}$ of the local updates, FedGSM uses a compensation term ${\bf\Delta}_n^{r_n(j)}$, which is equal to the difference between the local SGD progress (\ref{eq:local_update}) of two consecutive rounds, to combat staleness. This compensation term is added to the local updates. Since the satellite orbits are deterministic, the staleness effect for two consecutive rounds is comparable, and ${\bf\Delta}_n^{r_n(j)}$ can be taken as the accurate change of the local gradient, which can be used to correct the local update. Furthermore, ${\bf \bm{\nabla}}_n^i$ can be computed in the client side conveniently and sent back to the server without additional communication overhead.
FedGSM addresses the issue of staleness by introducing a compensation term ${\bf\Delta}_n^{r_n(j)}$ in place of directly sending the incremental $\Wbf_n^{r_n(j-1), E} - \Wbf_n^{r_n(j-1), 0}$ of the local update. The compensation term ${\bf\Delta}_n^{r_n(j)}$ is calculated as the difference between the local SGD progress (\ref{eq:local_update}) of two consecutive rounds. This approach takes advantage of the deterministic nature of satellite orbits, where the staleness effect for two consecutive rounds is comparable. Therefore, ${\bf\Delta}_n^{r_n(j)}$ can be used as a reliable measure of the accurate change of the local gradient, allowing it to correct the local update. Additionally, computing ${\bf \bm{\nabla}}_n^i$ on the client side reduces the communication overhead, as it can be conveniently transmitted back to the server.

\end{remark}

\section{Simulations} \label{sec:sim}
In this section, we provide a numerical comparison of the performance of FedGSM with the state-of-art algorithms. We also provide an analysis of the system parameter $E$.

\begin{figure}[t!]
\centering
% \includesvg[inkscapelatex=false,width=3.4in]{./Result_Plot/CIFAR10_IID_updated4.svg}
% \includesvg[inkscapelatex=false,width=3.4in]{./CIFAR10_IID_updated_15.svg}
\includegraphics[width=3.4in,keepaspectratio=false]{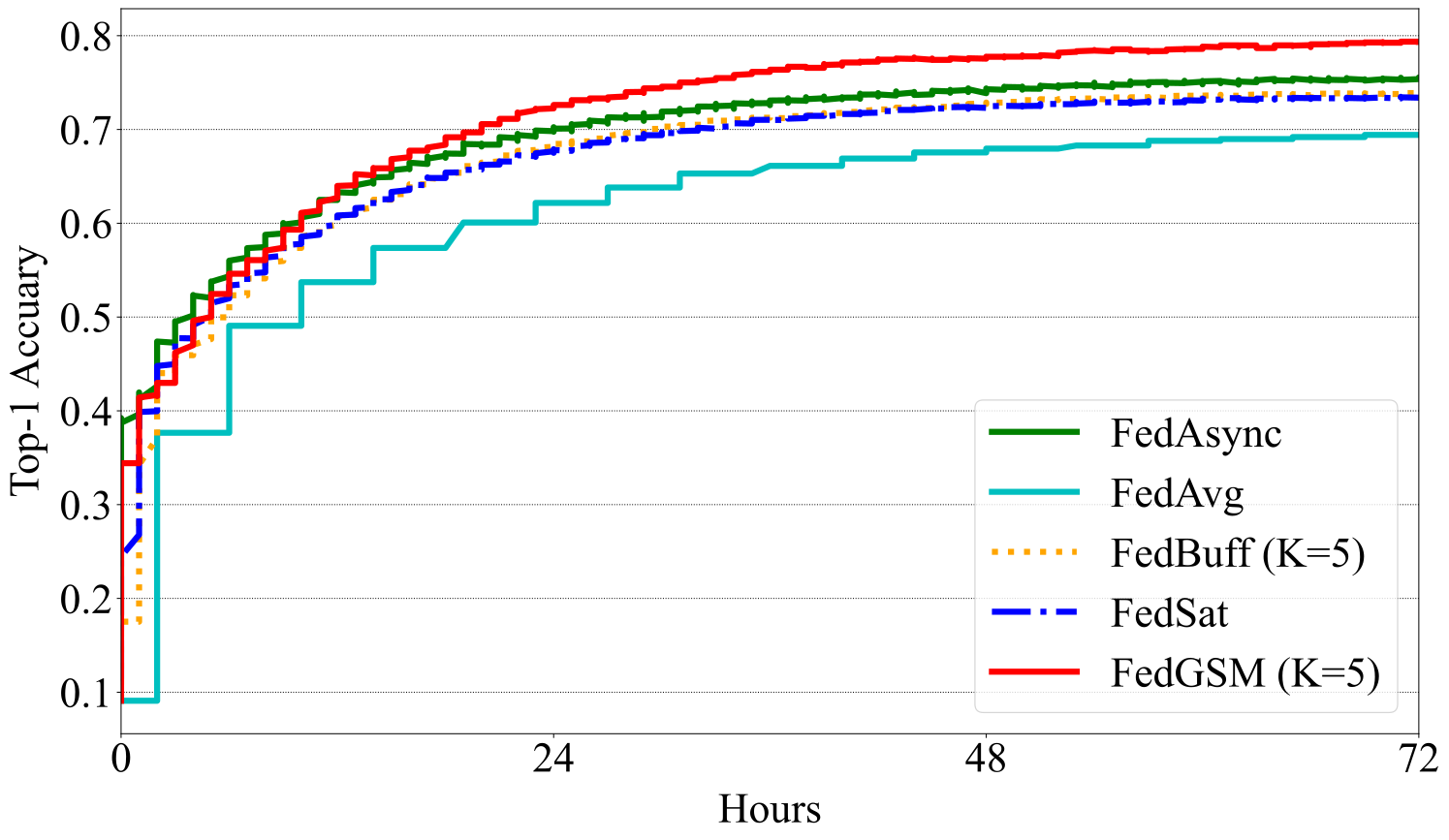}
\caption{Comparison of accuracy against observation time for FedGSM, FedAsync, FedAvg, FedBuff, and FedSat on Cifar10-IID with $E = 5$.}
\label{fig: Cifar10-IID}
\end{figure}

\textbf{Satellite Constellation and A Single GS.} We consider a Walker-delta constellation \cite{walker1984satellite} comprising $N = 40$ satellites distributed across $m = 10$ orbits with an inclination angle of $80\degree$. Among these, $5$ orbits are located at an altitude of $2000$ km, while the others are located at an altitude of $500$ km. Each orbit contains $4$ uniformly spaced satellites. The GS is situated at the North Pole and has a minimum elevation angle of $10\degree$. 
% Each simulation is run for a period of three days.

\textbf{Dataset, Model and Parameters.}
Our experiments use the CIFAR-10 dataset \cite{krizhevsky2009learning}, which is distributed to satellites in both IID and non-IID settings. In the IID setting, the samples are uniformly shuffled and equally assigned to each satellite. In the non-IID setting, the samples are distributed according to a Dirichlet distribution with a parameter of 0.3, as described in \cite{gao2022feddc}. The data volume across all satellites is uniform.

% The ML model is LeNet \cite{lecun1989handwritten}. Each satellite performs SGD with a initial local learning rate $\eta_{l} = 0.1$ and a decay rate $0.998$. The batch size is $10$ and the local epoch is $E = 5$. The global learning rate is $\eta_g = 0.1$ and the number of satellites participating an aggregation is $K=5$.

We adopt LeNet \cite{lecun1989handwritten} as our ML model, and each satellite performs SGD with an initial local learning rate of $\eta_l = 0.1$ and a decay rate of $0.998$. The batch size is set to $10$, and the local epoch is set to $E = 5$. The global learning rate is $\eta_g = 0.1$, and $K=5$ satellites participate in each aggregation round.

\begin{figure}[t!]
\centering
% \includesvg[inkscapelatex=false,width=3.4in]{./Result_Plot/CIFAR10_NONIID03_updated.svg}
% \includesvg[inkscapelatex=false,width=3.4in]{./Result_Plot/CIFAR10_NONIID03E5_updated4.svg}
% \includesvg[inkscapelatex=false,width=3.4in]{CIFAR10_NONIID03E5_updated_15.svg}
\includegraphics[width=3.4in,keepaspectratio=false]{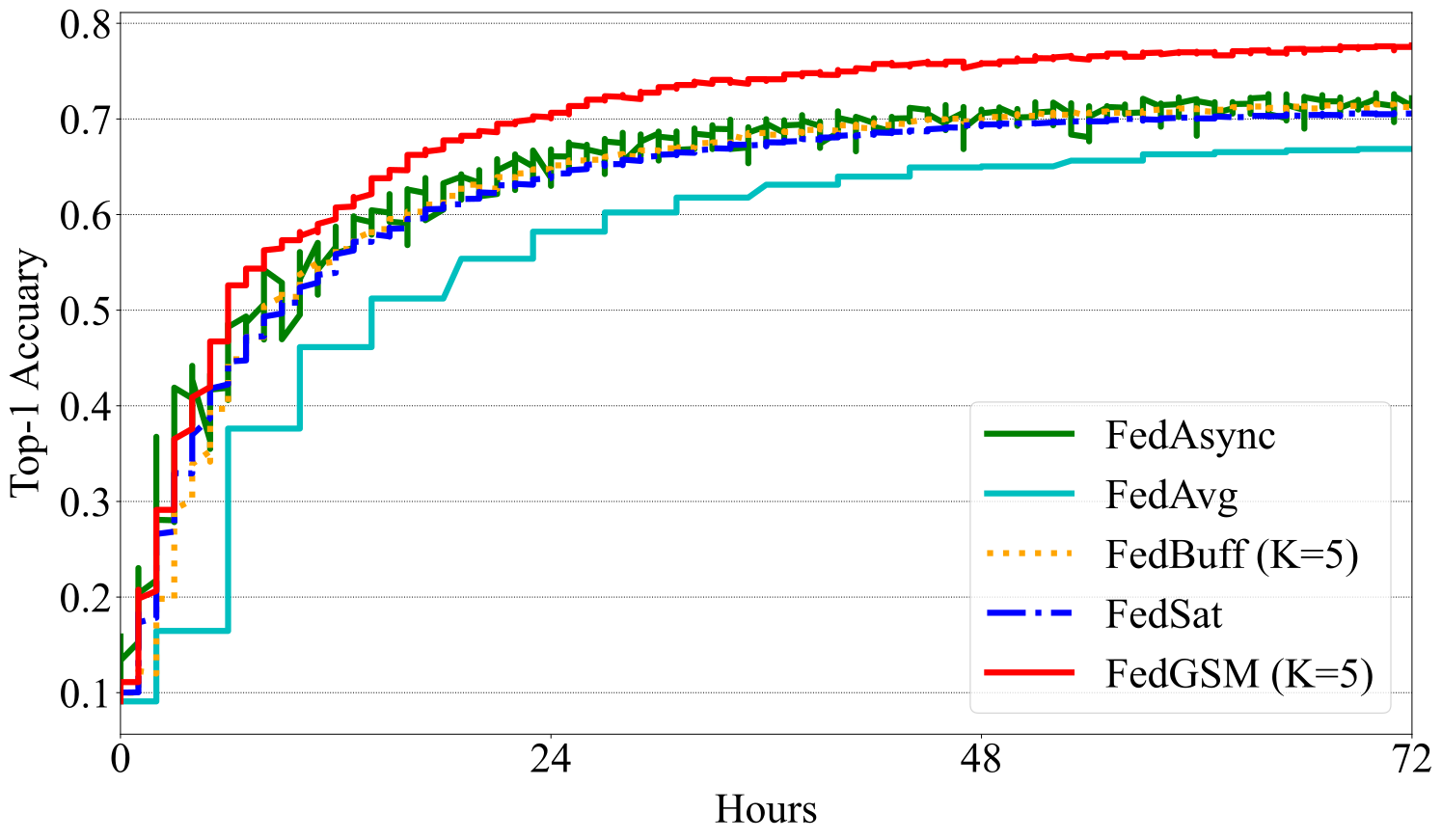}
\caption{Comparison of accuracy against observation time for FedGSM, FedAsync, FedAvg, FedBuff, and FedSat on Cifar10-nonIID (Dirichlet-0.3) with $E = 5$.}
\label{fig:Cifar10-nonIID(0.3)}
\end{figure}

\begin{figure}[t!]
\centering
% \includesvg[inkscapelatex=false,width=3.4in]{./Result_Plot/CIFAR10_NONIID03E2515_updated_15.svg}
% \includesvg[inkscapelatex=false,width=3.4in]{CIFAR10_NONIID03E2515_updated_20.svg}
\includegraphics[width=3.4in,keepaspectratio=false]{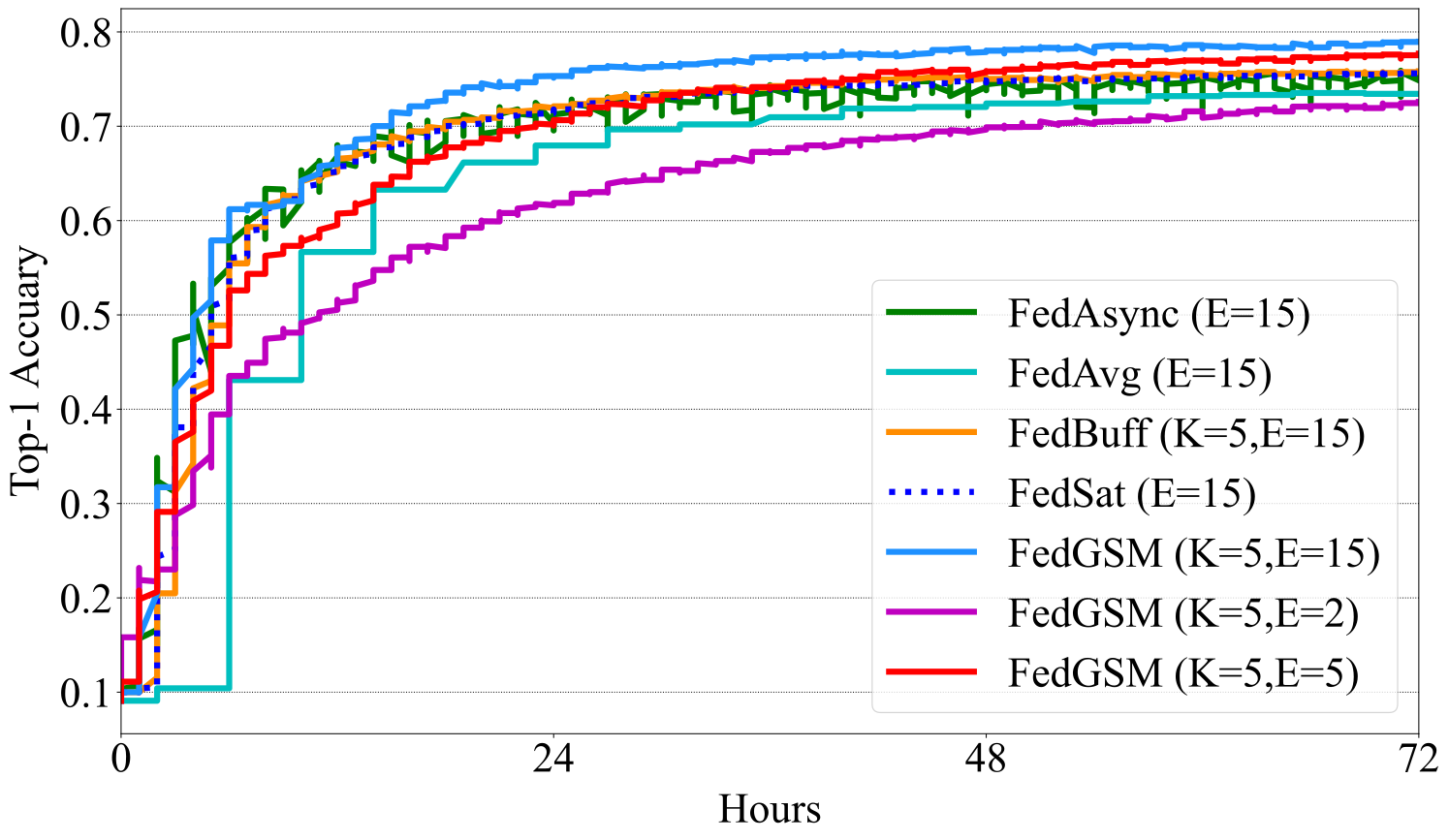}
\caption{Comparison of accuracy against observation time for FedGSM, FedAsync, FedAvg, FedBuff, and FedSat on Cifar10-nonIID (Dirichlet-0.3) with different values of $E$.}
\label{fig:Cifar10-nonIID(0.3)-E}
\end{figure}

\textbf{Baselines.}
We compare FedGSM with FedSat \cite{fedsat}, FedBuff \cite{fedbuff}, FedAsync \cite{fedasync}, and FedAvg \cite{FedAvg}. 
% FedBuff and FedAsync are universal asynchronous FL algorithms, while FedAvg is a classical synchronous FL algorithm. They are suitable for satellite communication scenarios.
The training process of FedBuff is the same as in Algorithm \ref{alg:FedGSM-SERVER} and Algorithm \ref{alg:FedGSM-Client}, except that the ${\bf \bm{\nabla}}_n^{r_n\left(j\text{+}1\right)}$ in line $7$ in Algorithm \ref{alg:FedGSM-Client} is substituted by ${\Wbf}_n^{r_n\left(j\right),E} - {\Wbf}_n^{r_n\left(j\right),0} $. 
The implementation of FedAvg and FedAsync in satellite communication scenarios is straightforward, with details that can also be found in \cite{fedsat}. 
% Specifically, the ${\bf \bm{\nabla}}_n^i$ in FedSat is substituted by ${\Wbf}_n^{i,E}-{\Wbf}_n^{old}$. FedSat is an asynchronous algorithm specifically proposed for satellite communication scenarios. Its aggregation strategy is the same as FedAvg employing the weighted average of $N$ local models as the global model, which only one client participating per round. 
FedSat is an asynchronous algorithm specifically proposed for satellite communication scenarios. The ${\bf \bm{\nabla}}_n^{r_n\left(j\text{+}1\right)}$ in FedSat is substituted by ${\Wbf}_n^{r_n\left(j\right),E}-{\Wbf}_n^{r_n\left(j\shortminus1\right),E}$. 

\textbf{Results Analysis.} Fig. \ref{fig: Cifar10-IID} and \ref{fig:Cifar10-nonIID(0.3)} show the test accuracy performance of five schemes with $E =5$ on two data settings. It is shown that FedAvg, a synchronous scheme, performs the worst by waiting for all the satellites to send the results back. Thanks to the compensation mechanism, FedGSM can achieve a higher test accuracy than all the other algorithms for both cases. This also demonstrates the negative effect of gradient staleness by simply using the standard update $\Wbf_n^{r_n(j-1), E} - \Wbf_n^{r_n(j-1), 0}$.

% For both cases, FedGSM outperforms the other algorithms by leveraging the compensation scheme, achieving a higher test accuracy. 

% FedGSM effectively compensates for the large staleness through global and local acceleration with historic knowledge, thus accomplishing a breakthrough in overall performance and achieving higher recognition accuracy.

% Fig. \ref{fig:Cifar10-nonIID(0.3)-E} shows the text accuracy performance of five schemes with different values of $E$. It is shown that FedGSM with $E=15$ can achieve a slighter higher accuracy than with $E=5$. In other words, more local iterative computations can accelerate the training process. This would not be the case as $E$ increases since the local updates might diverge especially for nonIID datasets. Overall, both choices outperform the other schemes. However, a small value of $E$ is not a good choice.

Fig. \ref{fig:Cifar10-nonIID(0.3)-E} presents the accuracy performance of five different schemes using varying values of $E$. The results indicate that FedGSM with $E=15$ achieves slightly higher accuracy compared to the case where $E=5$, indicating that more local iterative computations can expedite the training process. This also requires a higher computing capacity for LEO satellites. However, increasing $E$ beyond a certain threshold may result in divergence of local updates, particularly for non-IID datasets. Overall, both choices outperform the other schemes, but a small value of $E$ is not a recommended choice.

\section{Conclusions} \label{sec:con}

% This paper introduces FedGSM, an asynchronous FL algorithm designed to mitigate the issue of information staleness in satellite communication scenarios. FedGSM leverages the global information of model differences during aggregation and compensates for the staleness of local pseudo-gradients with local information of pseudo-gradient differences. Notably, these improvements are achieved without incurring additional communication or equipment deployment costs. Simulation results show that FedGSM outperforms baseline asynchronous algorithms, delivering impressive accuracy gains for both the IID and non-IID CIFAR-10 datasets. 

In this paper, we have presented FedGSM, an asynchronous FL algorithm designed to address the issue of gradient staleness in satellite communication scenarios. FedGSM uses the difference of the local model difference, which offsets the staleness effect, to compensate for the staleness of local models. Importantly, FedGSM achieves these improvements without incurring additional communication costs. Our simulation results demonstrate that FedGSM can enhance accuracy improvement on both the IID and non-IID CIFAR-10 datasets, as compared to the baseline algorithms. 

% We believe that FedGSM represents a valuable contribution to the field of FL and has the potential to improve the accuracy of machine learning models trained on distributed systems.


% \bibliographystyle{IEEEtran}
\bibliographystyle{IEEEtranN}
% \bibliography{cite.bbl}
\end{document}